# Blockchain based Decentralized Petition System

Dr. Jagdeep Kaur (Mentor)
*(Dr. B. R. Ambedkar NIT Jalandhar)*
Jalandhar, India
kaurj@nitj.ac.in

*Kevin Antony[1]*
*(Dr. B. R. Ambedkar NIT Jalandhar)*
Jalandhar, India
kevina.cs.20@nitj.ac.in

Nikhil Pujar[2]
*(Dr. B. R. Ambedkar NIT Jalandhar)*
Jalandhar, India
nikhilrp.cs.20@nitj.ac.in

Ankit Jha[3]
*(Dr. B. R. Ambedkar NIT Jalandhar)*
Jalandhar, India
ankitj.cs.20@nitj.ac.in

*Abstract*—This paper aims to review research papers written on petition systems. It also highlights the shortcomings of the existing petition systems and how our blockchain based implementation will overcome these shortcomings.

I. INTRODUCTION

II. Decentralised voting systems have piqued the curiosity of many people in recent years owing to its ability to provide safe and transparent voting platforms without the need of middlemen. Such systems can help address issues like voter fraud, manipulation, and trust in the electoral process. In this study, we present a decentralized voting system web application that employs blockchain technology to assure the voting process's integrity and security. Our proposed system aims to provide a transparent, decentralized decision-making process that enables every vote to be counted while eliminating the need for centralized authorities. We present an overview of the system architecture, design considerations, and implementation details, along with the potential benefits and limitations of the proposed system. Finally, we discuss future directions for research in this field.

III. Decentralized voting systems have gained increasing attention in recent years as a potential solution to the problems of traditional centralized voting systems, such as lack of transparency, security vulnerabilities, and susceptibility to manipulation. In this research paper, we present a web application for decentralized voting that employs blockchain technology and smart contracts to ensure fairness, transparency, and tamper-proofness. We also examine the technical aspects of the application, including the underlying algorithms and protocols, and discuss the potential benefits and challenges of decentralized voting systems. Our study intends to contribute to current efforts to improve the transparency and accessibility of democratic processes.

A. *Overview*

The process of creating and signing e-petition has the potential to be revolutionised by a blockchain-based e-petition app. Blockchain technology enables the creation of a trusted platform where people may develop and sign petitions. Due to the decentralised nature of each signature's recording and the consequent impossibility of data manipulation, the implementation of blockchain completely removes the potential of fraudulent operations. Additionally, the usage of smart contracts guarantees that the petition's provisions are automatically enforced, providing consumers with an additional degree of protection.

Additionally, real-time updates would be possible through the app, alerting users if a petition receives a particular amount of signatures or hits a significant milestone. Users may track the development of a petition and its influence on society thanks to this feature, which fosters a feeling of community. Additionally, the app may be connected to social networking sites, which would make it simpler to share and raise awareness of a specific cause or problem.

In conclusion, a blockchain-based e-petition software has the ability to develop a safe and open platform for users to generate and sign petitions. The use of smart contracts adds an additional degree of protection while ensuring that the data is secure and unchangeable thanks to the usage of blockchain technology. The social media integration and real-time updates of the app would promote a feeling of community and make it simpler to raise awareness of significant topics.

B. *Drawbacks of existing Petition Systems*
   1. Easily Tamperable

Online petitions are easily tamperable because they are typically hosted on a website or platform that can be accessed by anyone with an internet connection. This means that anyone with the right technical knowledge or tools can potentially manipulate the petition in various ways.

Here are some common ways that online petitions can be tampered with:

1. Automated Bots: Automated bots can be used to create fake signatures on the petition. These bots can be programmed to make it appear as if a large number of people have signed the petition, even though they are not real people.

Identify applicable funding agency here. If none, delete this text box.

XXX-X-XXXX-XXXX-X/XX/$XX.00 ©20XX IEEE

2. Multiple Signatures: It is easy for individuals to sign a petition multiple times by simply using different email addresses or IP addresses.

3. Fake Signatures: People can establish false email accounts or use someone else's email address to join the petition, giving the impression that the petition has more support than it does.

4. Manipulating Results: The results of the petition can be manipulated by individuals who have access to the back-end of the petition website. They can delete or add signatures, change the wording of the petition, or modify the results to suit their own agenda.

5. Hacking: The website hosting the petition can be hacked, and the data can be altered or stolen. This can affect the credibility of the petition and make it difficult to know if the signatures are genuine.

Overall, while online petitions can be a useful tool for mobilizing public support for a cause, they are not always reliable or accurate. It's important to be aware of the potential for tampering and to verify the authenticity of any petition before relying on it as evidence of public opinion.

2. Lack of verification: Online petitions can be easily manipulated, as there is no way to verify the identities of the signatories. This can lead to fraudulent signatures, which can undermine the credibility of the petition.

3. Over-reliance on technology: Online petitions depend on technology, which can be prone to glitches and errors. If the website hosting the petition goes down or experiences technical difficulties, it may prevent people from signing or accessing the petition.

4. Dilution of impact: The ease of creating and sharing online petitions has led to an oversaturation of petitions on various issues. This can dilute the impact of individual petitions, as people may become desensitized to the sheer number of petitions circulating online.

*C. Features*

A blockchain-based online petition system would have the following features:

1. Decentralization: The system would be decentralized, meaning that there would be no central authority controlling the petition platform. Instead, the system would be distributed across a network of computers.

2. Transparency: Since a blockchain is a transparent and immutable ledger, every signature and transaction on the platform would be publicly visible and verifiable.

3. Security: The blockchain's security measures, such as encryption and digital signatures, would ensure that every signature on the platform is authentic and cannot be tampered with.

4. Privacy: While the system would be transparent, the identities of individual signatories could be kept private, ensuring that they can safely express their views without fear of retribution.

5. Smart contracts: Smart contracts could be used to automate the verification process and ensure that petitions meet certain criteria before they are published on the platform.

6. Token economics: Tokens could be used to incentivize users to participate in the petition system, and to reward those who contribute to the success of a petition.

7. Community governance: A blockchain-based petition system could be governed by a community of users, who would make decisions about the platform's development and management.

Overall, a blockchain-based online petition system would provide a secure, transparent, and decentralized platform for people to express their views and push for change.

*D. Research Paper Abstracts*

- https://www.researchgate.net/publication/49611690_Electronic_Petitions_and_Institutional_Modernization_International_Parliamentary_E-Petition_Systems_in_Comparative_Perspective

  Petitioning political leaders has occurred since the beginning of time, but recent developments are the consequence of the internet's rise as a significant mass communication medium.Epetitions were first designed as official online engagement routes by governments and parliaments, with the British Prime Minister's e-petition system acting as the most well-known example.Online voting for general elections and referendums was just recently made available in Estonia, but online consultations are more common but are not yet technically regulated and are primarily voluntary.This article investigates the relationship between public institutions and internet-based platforms that aim to give additional and/or new opportunities for political activity.

It attempts to identify the essential institutional and technological features of the various e-petition platforms, as well as how and by whom they are used, in order to draw conclusions about how "offline" institutions, technology design, and political activity interact. The Norwegian local level, the Scottish Parliament, the German Bundestag, and the Queensland Parliament will all be investigated.

The second email poll showed that the passage of the signature barrier is crucial for e-petitions to encourage citizen engagement. Two ideal groups—one with limited influence and the other with strong influence—were made in order to derive conclusions. It was discovered that each group's average municipality received 302 e-petitions on average, of which 60 reached the required number of signatures and 10 were actually implemented.

Municipal websites' design elements were looked at to see how they affected influence. Four designs that were unique to the two groups were produced as a consequence of the data that was gathered. These were the "Hot" issue, the time limit, the social role, and the signature threshold. The 34 municipalities under study have significantly different signature requirements, ranging from 20 votes for Västervik to 500 votes for Värmdö. An e-petition has between 30 days and a year to gather a certain amount of signatures. Votes in the High Influence group had a 90-day time restriction, while votes in six other municipalities had a 60-day time limit and votes in three municipalities had to be gathered in 30 days.

- https://journals.sagepub.com/doi/full/10.1177/21582440211001354

This study sought to identify causality, also known as cause and effect, and explain the link between an independent and dependent variable. (Bryman, 2012). The e-petition's influence and design serve as the two independent variables that demonstrate this link. In order to collect the data necessary to address the two components of the research question, one approach that was divided into many phases was used. The goal was to identify the whole population of municipalities utilising e-petitions by sending an email to all 290 Swedish municipalities and requesting if they had implemented the e-petition. The emails were sent to the municipalities' official addresses, which were acquired via SALAR. (Swedish Association of Local Authorities and Regions).

The political system can be transformed by institutional and technological mediation, as discussed in both Anttiroiko (2003) and Sweden's second democracy report (2015). In this study, technology's applicability was evaluated along with the question, "Are the edemocracy experiments or practises such that people involved may truly influence the issues of interest?" to understand influence. In order to comprehend the impact of e-petition systems in 41 municipalities, the signature threshold of the e-petition was analysed. Questions included how many e-petitions had been received, how many had attained the signature threshold, and how many had obtained enough votes to be passed on. The second research topic was addressed using this information.

- https://www.sciencedirect.com/science/article/abs/pii/S0740624X21000058

The Internet has enabled new forms of participation and collective action, such as e-petitioning, which allows for the quick, easy, and accessible activation of large numbers of people. This might generate a sense of shared identity among loosely affiliated multinational lobbying organisation. Following the kidnapping of 276 young female students from the Government Ladies Secondary School in Chibok, Borno State, Nigeria, by heavily armed men in 2014, the "Bring Back Our Girls" petition was launched. It was later revealed that the girls were abducted by members of Boko Haram, an Islamic extremist organisation with origins in northeastern Nigeria, with the intention of selling them as slaves for marriage. Ify Elueze, a young Nigerian girl, initiated the petition "All World Leaders Bring Back Nigeria's 200 Missing School Girls #BringBackOurGirls" on Change.org. By May 7, 2014, the petition had gathered over 250,000 signatures, and by May 15, 2015, it had collected 1,104,440 signatures from individuals all around the world. Michelle Obama, the First Lady of the United States, utilised Twitter to raise awareness by posting a photo of herself holding a banner that said, "Bring Back Our Girls."

E-petitions are a sort of online collective action that collects signatures in order to be successful and maybe effective in influencing the responses of decision makers. Through the investigation of a case study, this article relies on two well-known agenda framing theories to analyse the mechanisms by which e-petitions accumulate signatures. Policy agenda setting theory focuses on the circumstances under which topics contend for attention and are acted upon by government decision makers, whereas media agenda setting theory focuses on how policy areas become viewed as relevant by the general public.

- https://www.diva-portal.org/smash/get/diva2:1429781/FULLTEXT01.pdf

With the introduction of digital technology into political processes in government, individuals are now able to weigh in on political matters such as legislative amendments or altered service and programme administration. Attempts towards e-democracy have been made in IS research, with an emphasis on the inventive and deliberative elements of emerging technologies. We must analyse digital technology in connection to the "Four Is of institutions, influence, integration, and interaction" in order to comprehend its significance in the transformation of political systems. Sweden uses a variety of digital technologies in its government, and the term "e-förvaltning" is frequently used to distinguish between its three main components: e-services, e-government, and edemocracy. As digital technologies proliferate in Sweden, there has been an imbalance in research on e-government, e-service, and e-democracy.

- https://firstmonday.org/ojs/index.php/fm/article/view/6001/5910

Using quantitative content analysis of 220 unofficial e-petitions on the website adressit.com, this study investigates anonymous political engagement in the form of e-petition signing. The results show that one of the most important predictors of the percentage of anonymous signatures is the kind of demand. Anonymity and democracy coexist in an ambivalent way, with voting secured by secret ballots and significant campaign contributions being made public. We may learn more about citizens' political behaviour and the variables influencing people's decisions to participate in politics anonymously by examining the instances when people opt to keep their signatures from public inspection. This study aims to deepen understanding of the trends behind anonymous e-petition signing. The purpose of the study is to look into the association between e-petition attributes and citizen disclosure of signatures in a representative democracy. Which e-petition features affect the proportion of anonymous signatories?

- https://www.tandfonline.com/doi/abs/10.1080/20421338.2020.1835174

The goal of this survey was to ascertain how City of Tshwane, South Africa inhabitants felt about the electronic petition system. Results indicate that citizens are prepared for and belief in the efficacy of the e-petition system. It comes to the conclusion that when utilised in conjunction with the conventional paper petition method, the e-petition system may be successful. Additionally, it offers policymakers, municipal management, and government insights and gives recommendations for further research.

- https://www.iprojectmaster.com/computer-science/final-year-project-materials/design-and-implementation-of-an-online-petition-management-system

A number of institutions, including legislatures and government agencies, have implemented electronic petitions systems. (e-petitions). It has been possible to petition political leaders or other governmental agencies since antiquity. (Bockhofer 1999; Hirsch 2007; Klasen 1991). This route of communication between subjects and rulers was constantly modified over the ages to meet the demands of shifting political and social conditions. E-petitions have advanced past the test organisation and are characterised by a substantial level of institutionalisation and procedural development in comparison to the majority of other petitioning methods made available by public entities. In order to improve the link between Internet-based petitioning and public institutions, it is suggested that this research create an e-petitioning system. When completely implemented, this will make it easier for members of the general public to engage with and petition governmental entities via various online platforms. The introduction of the online petition management system will facilitate thousands of handwritten petitions sent throughout the world, promote accountability, and organise them. PHP, mySQL, HTML, and CSS were used to create the suggested system. The major goal of this project is to build and construct an online petition management system that would allow citizens in democracies to submit electronic petitions to assist institutions and the government in making better decisions. Enable the electronic submission of petitions over the internet. Given that the government receives hundreds of petitions every day, they can be simply monitored. Boost accountability, improve top government institutions' governance, and advance the government's legislative objectives.

- https://link.springer.com/article/10.1007/s12394-009-0012-8

People sign petitions, which are formal requests made to an individual in authority. Citizens may express their support or disagreement with government policies and offer feedback to government agencies through petitions. In the real world, petition signers usually include a unique identity with their handwritten signature in order to filter out phoney or duplicate signatures (such as the national ID number).

- https://eprints.whiterose.ac.uk/137839/9/PaperFinalWithFigures.pdf

In order to significantly strengthen its public participation, the UK Parliament created an e-petition system in 2015. We investigate whether this aim is being met by studying Twitter data from conversations regarding petitions being debated in Parliament. We analyse what Twitter data indicates about the amount of engagement, the themes covered in Twitter e-petition dialogues, and the participants and interactions using natural language processing, machine learning, and social network analysis. Our research offers intriguing perceptions of how people view the fairness and responsiveness of e-petition processes, and it suggests that the objectives of the original petitions should be included more fully in parliamentary deliberations. The findings indicate that there were homophilic tendencies in the Twitter e-petition talks. Last but not least, our study demonstrates that Twitter conversations on e-petitions frequently occur inside comparable networks and that homophily predominates in these talks. Although homophily was discovered to be considerable, it was not adequate to draw the conclusion that the Twitter discussion network was polarised since some degree of homophily is anticipated in political disputes and is only an indicator of displays of solidarity and mobilisation between activists.

Additionally, the mild homophily demonstrates that there was some communication between opposing political factions.

Our research also highlights the need of big data analysis to comprehend contemporary technologies like e-petitions since they let us track public sentiment in real-time

- https://www.researchgate.net/publication/300415479_Success_Factors_of_Online_Petitions_Evidence_from_Changeorg

Using data from Change.org, a major social advocacy website, this research paper explores the success characteristics of online petitions. We uncover the primary determinants of successful petitions using a quantitative study of over 12,000 petitions, including criteria such as the petition topic, the quantity and kind of supporters, and the inclusion of multimedia material. The findings provide light on the mechanics of online activism and propose tactics for petition producers to boost their chances of success. Furthermore, the study adds to the expanding body of research on the influence of digital technology on political engagement and social transformation.

- https://journals.sagepub.com/doi/10.1177/21582440211001354

This research paper proposes an assemblage-based framework for the analysis and research of e-petitions, aiming to provide a new theoretical lens for understanding the complex socio-technical interactions and power relations involved in online activism. Drawing on a review of the literature and case studies of e-petitions in the UK and Greece, the paper illustrates how the framework can be applied to investigate the formation, functioning, and effects of e-petition assemblages. The study highlights the significance of material and discursive elements, as well as the role of human and non-human actors, in shaping e-petition movements. Overall, the work advances our understanding of the dynamics of online activism and serves as a significant resource for scholars and practitioners.

- https://www.pewresearch.org/internet/2016/12/28/we-the-people-five-years-of-online-petitions/

This research report by the Pew Research Center examines the evolution and impact of online petitions in the United States over the past five years. Using a mix of quantitative and qualitative methods, the study analyzes data from the White House's We the People petition platform and interviews with petition creators and supporters. The report provides insights into the demographics, topics, and outcomes of online petitions, as well as the motivations and experiences of those who participate in them. The study also explores the challenges and opportunities of online activism in the context of democratic engagement and civic discourse. Overall, the research helps us comprehend the role of digital technology in affecting political engagement and societal transformation.

- https://journals.plos.org/plosone/article?id=10.1371/journal.pone.0178062

This research paper investigates the temporal dynamics of online petitions using data from Change.org and Avaaz. The study explores how the number and rate of signatures vary over time and identifies patterns of growth, decline, and resurgences in petition activity. The paper also examines the impact of external events and media coverage on petition dynamics. Through a combination of statistical analysis and network modeling, the study reveals the complex interplay between human behavior and digital platforms in shaping online activism. Overall, the research advances our knowledge of the dynamics of collective action in the digital age and has consequences for the design and administration of online petition systems.

- https://ieeexplore.ieee.org/document/9530852

This study proposes a model of online petition signing dynamics using Taiwan's Join Platform. The study utilizes data from over 500 petitions and 400,000 signatures to examine the factors that influence the growth and decline of online petitions. Using a combination of statistical analysis and network modeling, the authors identify several key predictors of petition success, including the number and diversity of supporters, the use of persuasive messages, and the influence of external events. The paper also highlights the importance of context-specific factors in shaping online activism, such as cultural norms and political structures. Overall, the study adds to the increasing body of literature on internet activism and sheds light on the dynamics of collective action in Taiwan.

## Acknowledgment

We would like to take this opportunity to Dr Jagdeep Kaur for guiding us and Dr .B R Ambedkar NIT Jalandhar for providing a rich research culture.